\documentclass[a4paper,12pt]{iopart}
\usepackage[utf8]{inputenc}
\usepackage{hyperref}
\usepackage{color}

\newcommand{\eqend}[1]{\,\mathrm{#1}}

\newcommand{\bx}{\textbf{x}}

\newcommand{\bp}{\textbf{p}}

\newcommand{\p}{^{\prime}}

\newcommand{\laplace}{\bigtriangleup}

\makeatletter

\newcommand\fulllabel[1]{
\def\@currentlabel{\ifnumbysec\arabic{section}.\arabic{eqnval}\else\arabic{eqnval}\fi}
\label{#1}%
\let\@currentlabel\theequation
}

\let\@@ppendixstar\@appendixstar
\def\@appendixstar{\@@ppendixstar
\def\numparts{\addtocounter{equation}{1}%
     \setcounter{eqnval}{\value{equation}}%
     \setcounter{equation}{0}%
     \def\theequation{\ifnumbysec
     \Alph{section}.\arabic{eqnval}{\it\alph{equation}}%
     \else\Alph{section}\arabic{eqnval}{\it\alph{equation}}\fi}}
\def\endnumparts{\def\theequation{\ifnumbysec
     \Alph{section}.\arabic{equation}\else
     \Alph{equation}\fi}%
     \setcounter{equation}{\value{eqnval}}}
}

\makeatother

\usepackage{iopams}

\begin{document}

\title[IR divergences in cosmological spacetimes]{Infrared divergences for free quantum fields in cosmological spacetimes}

\author{Atsushi Higuchi\(^1\) and Nicola Rendell\(^1\)}
\address{\(^1\) Department of Mathematics, University of York, Heslington, York, YO10 5DD, United Kingdom}

\eads{\mailto{atsushi.higuchi@york.ac.uk}, \mailto{nlr512@york.ac.uk} }

\begin{abstract}
We investigate the nature of infrared divergences for the free graviton and inflaton two-point functions
in flat Friedman-Lema\^{\i}tre-Robertson-Walker spacetime.  These divergences arise because the momentum
integral for these two-point functions diverges in the infrared.
It is straightforward to see that the power of the momentum in the integrand
can be increased by $2$ in the infrared using large gauge transformations, which are sufficient for rendering these
two-point functions infrared finite for slow-roll inflation.  In other words, if the integrand of the momentum integral
for these two-point functions behaves like $p^{-2\nu}$,
where $p$ is the momentum, in the infrared, then it can be made to behave like $p^{-2\nu+2}$ by large gauge
transformations.  On the other hand, it is known that, if one smears these two-point
functions in a gauge-invariant manner,  the power of the momentum in the integrand is changed from
$p^{-2\nu}$ to $p^{-2\nu+4}$.  This fact
suggests that the power of the momentum in the integrand for these two-point functions can be increased
by $4$ using
large gauge transformations.  In this paper we show that this is indeed the case.   Thus,
the two-point functions for the graviton and inflaton fields can be made finite by large gauge transformations for
a large class of potentials and states in single-field inflation.
\end{abstract}

\submitto{\CQG}

\maketitle

\section{Introduction}

Our universe is believed to have experienced an inflationary period in its early stages of development. Inflation as a
theory of the early universe was proposed in \cite{InflationG,InflationS,InflationL,InflationAS} as a
solution to the flatness and horizon problems, among others, of the standard Big Bang model.
The background spacetime for the inflationary model is a spatially flat
Friedman-Lema\^{\i}tre-Robertson-Walker (FLRW)
spacetime that expands exponentially.  If the expansion is exactly exponential, the spacetime is de~Sitter space.

It is well known that the massless minimally-coupled scalar field in de~Sitter space and other
spatially flat FLRW spacetimes has a two-point function which is divergent in the infrared (IR)~\cite{FordParker2}.
It is also known that
the graviton field satisfies the same equation as the massless minimally coupled scalar field in FLRW spacetime.
Hence, if the scale factor and the state are such that the massless minimally coupled scalar field has an IR-divergent two-point
function, then the graviton field will have one also~\cite{FordParker}.
For the single-field inflationary model the two-point functions for the scalar and tensor
perturbations are IR-divergent in a similar manner.  The physical significance of these divergences
has been studied over the past several years (see,
e.g.~\cite{an_mott,gar_tana,tsa_wood_2008,jan_prok,jan_prok_2,rio_slo,bur_et,raj_ku_le,see,ur_ta_10,ka_mi_wo,kita2_1,kita2_2,gid_slo_1, ger_he_ta,gid_slo_2,gar_rig,xue_gao_bran,pim_sen_zal,sen_zal,as_bau_gr,ur_ta_loop,ta_ur_grav,fer_san_slo}).

As for the IR divergences for the tensor perturbations, i.e.\ the gravitons, in de~Sitter space
the IR-divergent piece of the two-point function can be written in pure-gauge
form~\cite{Allen:1986dd,Explicit2}.  It was noted that these divergences can be gauged away by linear
gauge transformations that correspond to global shear transformations~\cite{IRdivpaper}.
We note that
there is some debate over the use of these `large' gauge transformations~\cite{bound,Gaugingaway}, i.e.\
gauge transformations that do not become identity at spatial infinity.
However, as is argued in~\cite{IRdivpaper}, it is legitimate to use
large non-compactly supported gauge transformations if one is interested only in local physics. Briefly, this is because a large gauge
transformation can mathematically be made to be compactly supported, without changing local physics,
by multiplication with a smooth compactly supported function which is $1$ in the local region of interest and turned off
smoothly outside. Then the two-point function will be IR finite if the two points are in the region where this compactly supported function is $1$,
which is the region of interest, though it is IR divergent elsewhere.\footnote{Here we apply a specific compactly supported gauge
transformation to each mode function rather the corresponding momentum component of a single gauge transformation.}
In \cite{bound} the Aharonov-Bohm effect is listed as an example where gauge-dependent quantities might play
a role.  However, since the IR divergences in the graviton (or inflaton) two-point function are not a topological effect, this
does not serve as a good example for arguing
against using large gauge transformations to gauge away IR divergences.
We also point out that the distribution of gravitational
fluctuations in momentum space is unchanged by these large gauge transformations; only the
mode function for each value of the momentum is modified.

Mathematically, these IR divergences in the two-point function arise because the power of the momentum \( p\)
for small \(p\) in the integrand of the $p$-integral is negative and too large for it to converge.
We point out in this paper that global shear transformations and dilation render the two-point functions IR finite
for the tensor and scalar perturbations in slow-roll single-field inflation by increasing the power
of \( p\) in the IR by $2$, from \( p^{-2\nu}\), say, to \(p^{-2\nu+2}\).  These large gauge
transformations have been studied in relation to the so-called consistency relations in inflationary
cosmology~\cite{Maldacena:2002vr,Creminelli:2004yq,Assassi:2012zq,Biagetti:2017viz,Tanaka:2017nff}.
It is possible that the fact mentioned above is known in the cosmology community in some form, but it does not
seem to have been pointed out explicitly.

Now, it was shown recently that,
if one smears the IR-divergent graviton and inflaton two-point functions in a gauge-invariant manner, then the power of \( p \)
mentioned above is changed from $p^{-2\nu}$ to
$p^{-2\nu+4}$~\cite{Quant2,InflationGTpre}.
This suggests that there should be
 large gauge transformations that change the small-\(p\) behaviour of the mode functions of the graviton and inflaton
from $p^{-\nu}$ to $p^{-\nu+2}$ so that it changes in the integrands of
the two-point functions from \(p^{-2\nu}\) to \(p^{-2\nu+4}\).
In this paper we find such gauge
transformations.  We discuss these gauge transformations first for the tensor perturbations, as they are
universal for any FLRW spacetime.  We then discuss the case of single-field inflation with the emphasis
on the scalar perturbations.

The rest of the paper is organised as follows. In section~\ref{tensor-perturbation}
we set up a framework for the tensor perturbations about a background FLRW
metric. We provide a review of how these perturbations are quantised in section~\ref{quantisation}.
In section~\ref{gauge-transformation} we present large gauge transformations which increase the power of \( p\) in
the IR for the tensor mode functions by $2$, thereby increasing the power of \(p\) in the integrand of the graviton two-point
function by \( 4\).  We apply our analysis to FLRW spacetime with constant slow-roll parameter \( \epsilon \), related to the deceleration
parameter \(q\) by \(q= -1+\epsilon\), and
determine its value for which the IR divergences can be gauged away for the scale-invariant vacuum state.
In section~\ref{single-field} we review the quantisation of the scalar perturbations through the Sasaki-Mukhanov
variable. In section~\ref{IR-in-single-field} we present large gauge transformations which increase the power
of \( p\) in the two-point functions for both the scalar and tensor perturbations by \(4\).  Finally, in
section~\ref{discussion} we summarise and discuss our results.  Some technical results are provided in
\ref{Appendix-A} and \ref{Appendix-B}.

\section{Tensor perturbations of the FLRW metric}\label{tensor-perturbation}
We consider the gravitational tensor perturbations around a background FLRW metric in $n$ dimensions.   We let $n\geqslant 4$ throughout this paper.
For definiteness we assume that
the matter consists of a perfect fluid.
There are several actions proposed to describe a perfect fluid in general relativity~\cite{taub,schutz,brown}.  The action
due to Schutz~\cite{schutz} is
\begin{equation}
I = \int d^n x\, \mathcal{L} \eqend{,}
\end{equation}
where
\begin{equation}
\mathcal{L} = \frac{1}{\kappa^2} \sqrt{-g}\,R + \sqrt{-g}\, p(\mu,S) \eqend{.} \label{lagrangian}
\end{equation}
The quantity $S$ is called the specific entropy, and $\mu$ and $V_\mu$ are defined by
\begin{eqnarray}
\mu & = & \sqrt{-g^{\mu\nu}V_\mu V_\nu} \eqend{,} \label{mu-g}\\
V_\mu & = & \nabla_\mu \phi + \alpha\nabla_\mu \beta + \theta\nabla_\mu S\eqend{.}
\end{eqnarray}
The positive constant $\kappa$ is related to Newton's constant $G_N$ by \(\kappa^2 = 16 \pi G_N\).  The independent
variables are $g^{\mu\nu}$,
$\alpha$, $\beta$, $\phi$, $\theta$ and $S$.
The most relevant fact here is that the pressure $p$ depends on the metric through
equation~(\ref{mu-g}).  By using the relation~\cite{schutz}
\begin{equation}
\frac{\partial p}{\partial\mu} = \frac{\rho + p}{\mu}\eqend{,}
\end{equation}
where $\rho$ is the energy density,
one readily finds the standard Einstein equations with a perfect fluid:
\begin{equation}
R_{\mu\nu} - \frac{1}{2}g_{\mu\nu} R = \frac{\kappa^2}{2}\left[ (\rho + p)u_\mu u_\nu + p g_{\mu\nu}\right]\eqend{,} \label{FLRW-eq}
\end{equation}
where
\begin{equation}
u_\mu = \mu^{-1} V_\mu\eqend{,}
\end{equation}
where the perfect fluid model is used to model the radiation and matter phases of the early universe.

As is well known, the metric of the form $g_{\mu\nu} = a^2(\eta)\eta_{\mu\nu}$, where $\eta_{\mu\nu}$
is the metric of flat spacetime, is a solution of  (\ref{FLRW-eq}) if
$u_\mu  = t_\mu$, where
\begin{equation}
t^\mu = a(\eta)\left(\frac{\partial\ }{\partial\eta}\right)^\mu\eqend{,}
\end{equation}
and if
\begin{eqnarray}
\kappa^2 \rho & = & (n-1)(n-2)\left( \frac{a'}{a^2}\right)^2\eqend{,}  \label{rho}\\
\kappa^2 p & = & - 2(n-2)\frac{a''}{a^3} - (n-2)(n-5)\left( \frac{a'}{a^2}\right)^2\eqend{.} \label{p}
\end{eqnarray}
We consider the tensor perturbation $h_{\mu\nu}$ of the metric of the form,
\begin{equation}
g_{\mu\nu} = a^2(\eta)(\eta_{\mu\nu} +  h_{\mu\nu})\eqend{,}
\end{equation}
that is synchronous, transverse, and traceless.  That is, we require
that $h_{\mu\nu}$ have no component in the direction of $u^\mu$, i.e.\ $h_{0\mu} = 0$, and that
its spatial component be transverse, $\partial^j h_{ij} = 0$, and traceless, $\delta^{ij}h_{ij} = 0$, where $\partial_j$ is the spatial
derivative operator in flat space, and where the index is raised by Kronecker's delta, $\delta^{ij}$.  We write the space components of $h_{\mu\nu}$
after choosing this gauge as $h_{ij} = H_{ij}$.   From (\ref{FLRW-eq}) it is clear that the perturbations
described by $H_{ij}$
do not mix with perturbations of any other fields at first order.  We find that $H_{ij}$ satisfies the following equation to first order~\cite{FordParker}:
\begin{eqnarray} \label{fieldequation}
\frac{1}{a^n} \frac{\partial}{\partial \eta} \left( a^{n-2} \frac{\partial}{\partial \eta} H_{ij} \right) - \frac{1}{a^2} \triangle H_{ij} = 0 \eqend{,}
\end{eqnarray}
where $\triangle = \delta^{ij}\partial_i\partial_j$ is the Laplacian on flat space.

\section{Quantisation of the tensor perturbation} \label{quantisation}
In order to quantise the field $H_{ij}$ representing the tensor perturbations,
we first expand the Lagrangian~(\ref{lagrangian}) to second order in
$\tilde{h}_{\mu\nu} = a^2 h_{\mu\nu}$
with the conditions $\nabla_\mu \tilde{h}^{\mu\nu} = 0$ and ${\tilde{h}^\mu}_{\ \mu} = 0$.  Thus,
we find (up to a total derivative) the quadratic Lagrangian relevant to the
tensor perturbations is as follows:
\begin{eqnarray}
\mathcal{L}_{T} &  = &  \frac{\sqrt{-g}}{2\kappa^2}\left[   - \frac{1}{2}\nabla_\mu \tilde{h}_{\nu\lambda}\nabla^\mu \tilde{h}^{\nu\lambda}
- \frac{\kappa^2}{2(n-2)}(\rho - p)\tilde{h}^{\mu\nu}\tilde{h}_{\mu\nu} \right. \nonumber \\
&& \ \ \ \ \ \ \ \ \left. +   R_{\mu\nu}\tilde{h}^{\mu\alpha}{\tilde{h}^\nu}_\alpha
 + R^{\beta\delta \alpha\gamma}\tilde{h}_{\alpha\beta}\tilde{h}_{\gamma\delta}\right]\eqend{,} \label{simplified}
\end{eqnarray}
where $\rho$, $p$, $R_{\mu\nu}$ and $R^{\beta\delta\alpha\gamma}$ are the background quantities.  By substituting (\ref{rho}), (\ref{p}), and the formula
\begin{eqnarray}
R_{\mu\nu\rho\sigma} & = & H^2\left[ g_{\mu\rho} g_{\nu\sigma} - g_{\mu\sigma}g_{\nu\rho}
+ \epsilon (t_{\mu}t_{\rho}g_{\nu\sigma} + t_{\nu}t_{\sigma}g_{\mu\rho} - t_{\mu}t_{\sigma}g_{\nu\rho} - t_{\nu}t_{\rho} g_{\mu\sigma})\right]
\eqend{,}\nonumber \\
\end{eqnarray}
where
\begin{eqnarray}
H & = & \frac{a'}{a^2}\eqend{,}\label{def-H}\\
\epsilon & = & - \frac{H'}{H^2 a}\eqend{,} \label{def-epsilon}
\end{eqnarray}
we can simplify $\mathcal{L}_T$ as
\begin{equation}
\mathcal{L}_T = \frac{1}{4\kappa^2} a^{n-2}\left( H'_{ij}H^{ij\prime} + H_{ij} \laplace H^{ij}\right)\eqend{,}
\end{equation}
where $H_{ij}'$ is the partial derivative of $H_{ij}$
with respect to $\eta$ and where $\laplace = \partial_k \partial_k$.


The quantisation of the field $H_{ij}(\eta,\mathbf{x})$ is standard.  It is expanded in terms of the mode functions 
$\gamma_{ij}^{(s,\mathbf{p})}(\eta,\mathbf{x})$
and their complex conjugates as
\begin{equation}
H_{ij}(\eta,\mathbf{x}) = \int \frac{\mathrm{d}^{n-1}\mathbf{p}}{(2\pi)^{n-1}}
\sum_s \left[ a_s(\mathbf{p})\gamma_{ij}^{(s,\mathbf{p})}(\eta,\mathbf{x})
+ a_s^\dagger(\mathbf{p})\gamma_{ij}^{(s,\mathbf{p})*}(\eta,\mathbf{x})\right]\eqend{.}
\end{equation}
The mode functions $\gamma_{ij}^{(s,\mathbf{p})}(\eta,\mathbf{x})$ are given by
\begin{equation} \label{tensor-mode}
\gamma^{(s,\mathbf{p})}_{ij}(\eta,\mathbf{x}) = \epsilon_{ij}^{(s)}(\mathbf{p})f_p (\eta)\mathrm{e}^{\mathrm{i}\mathbf{p}\cdot\mathbf{x}}
\eqend{,}
\end{equation}
where the polarisation tensors $\epsilon_{ij}^{(s)}(\mathbf{p})$ are traceless and satisfy 
$\epsilon_{ij}^{(s)}(\mathbf{p})p^j = 0$ and
\begin{equation}\label{polarisation-tensor-normalise}
\sum_{ij} \epsilon^{(s)}_{ij}(\mathbf{p}) \epsilon^{(r)}_{ij}(\mathbf{p}) = \delta^{sr}\eqend{.}
\end{equation}
The functions $f_p(\eta)$, where $p=|\mathbf{p}|$, satisfy
\begin{equation}\label{eq-mo}
\frac{1}{a^{n-2}(\eta)}\frac{\mathrm{d}\ }{\mathrm{d}\eta}\left[ a^{n-2}(\eta)\frac{\mathrm{d}\ }{\mathrm{d}\eta}f_p(\eta) \right]
+ p^2 f_p(\eta) = 0\eqend{,}
\end{equation}
which, of course, agrees with (\ref{fieldequation}).
Since the equation of motion (\ref{eq-mo}) implies
\begin{equation}
\frac{\mathrm{d}\ }{\mathrm{d}\eta}\left\{ a^{n-2}(\eta)
\left[ f^*_p(\eta) \frac{\mathrm{d}f_p(\eta)}{\mathrm{d}\eta}
- \frac{\mathrm{d}f^*_p(\eta)}{\mathrm{d}\eta} f_p(\eta)\right] \right\} = 0\eqend{,}
\end{equation}
it is possible to choose the normalisation of $f_p(\eta)$ such that
\begin{equation}\label{Wronskian}
f^*_p(\eta) \frac{\mathrm{d}f_p(\eta)}{\mathrm{d}\eta}
- \frac{\mathrm{d}f^*_p(\eta)}{\mathrm{d}\eta} f_p(\eta) = - \frac{\mathrm{2\mathrm{i}\kappa^2}}{a^{n-2}(\eta)}\eqend{.}
\end{equation}
With this choice we find
\begin{equation}\label{creation-annihilation}
[ a_s(\mathbf{p}), a_{s'}^\dagger (\mathbf{p}') ] = (2\pi)^{n-1}\delta_{ss'}\delta^{n-1}(\mathbf{p}-\mathbf{p}')\eqend{.}
\end{equation}

One defines the vacuum state $|0\rangle$ by requiring that $a_s(\mathbf{p})|0\rangle=0$ for all $s$ and $\mathbf{p}$.
Thus, the choice of
the function $f_p(\eta)$ satisfying (\ref{Wronskian}) determines the vacuum state. The two-point correlation function for
$H_{ij}(\eta,\mathbf{x})$
can be found using (\ref{creation-annihilation}) as
\begin{eqnarray}
\Delta_{iji'j'}(\eta,\mathbf{x};\eta',\mathbf{x}')
& : = & \langle 0|H_{ij}(\eta,\mathbf{x})H_{i'j'}(\eta',\mathbf{x}')|0\rangle \nonumber \\
& = & \int \frac{\mathrm{d}^{n-1}\mathbf{p}}{(2\pi)^{n-1}}
\sum_s \gamma_{ij}^{(s,\mathbf{p})}(\eta,\mathbf{x})\gamma_{i'j'}^{(s,\mathbf{p})*}(\eta',\mathbf{x}')\eqend{.}
\label{two-point}
\end{eqnarray}

It will be useful for later purposes to examine the solution $f_p(\eta)$ for small $p$. If $p=0$, two independent real solutions
$f_0(\eta) = F^{(1)}_0(\eta)$ and $F^{(2)}_0(\eta)$, can be chosen as
\begin{eqnarray}
F^{(1)}_0(\eta) & = & 1\eqend{,}\\
F^{(2)}_0(\eta) & = & \int  \frac{\mathrm{d}\eta}{a^{n-2}(\eta)}\eqend{,} \label{def-F2}
\end{eqnarray}
where the constant of integration is chosen suitably in (\ref{def-F2}).
Two independent real solutions, $F^{(1)}_p(\eta)$ and $F^{(2)}_p(\eta)$, can be found such that
\begin{equation}\label{F-Wronskian}
F^{(1)}_p(\eta)\frac{\mathrm{d}F^{(2)}_p(\eta)}{\mathrm{d}\eta} -
\frac{\mathrm{d}F_p^{(1)}(\eta)}{\mathrm{d}\eta} F_p^{(2)}(\eta) = \frac{1}{a^{n-2}(\eta)}\eqend{,}
\end{equation}
and that
\begin{equation}\label{FI}
F^{(I)}_p(\eta) = F^{(I)}_0(\eta) + \mathcal{O}(p^2)\eqend{,}\ I=1,2\eqend{.}
\end{equation}
This is because the $p$-dependence in (\ref{eq-mo}) is through $p^2$.  The solutions $f_p(\eta)$ can be expressed as
\begin{equation}\label{convention}
f_p(\eta) = \mathrm{i}A^{(T)}(p)F^{(1)}_p(\eta) + B^{(T)}(p)F_p^{(2)}(\eta) \eqend{.}
\end{equation}

The functions $F^{(1)}_p(\eta)$ and $F^{(2)}_p(\eta)$ are finite in the limit $p\to 0$, and the source of IR singularities
is the singular behaviour of $A^{(T)}(p)$ in this limit.
After choosing the real solutions $F^{(1)}(\eta)$ and $F^{(2)}(\eta)$  it is always possible to choose
$A^{(T)}(p)$ and $B^{(T)}(p)$ to be real.  This is done by
first choosing $B^{(T)}(p)$ to be real with adjustment of the phase factor,
and then absorbing any imaginary part of $A^{(T)}(p)$ with the redefinition of
$F^{(2)}(\eta) - [\mathrm{Im}\,A^{(T)}(p)/B^{(T)}(p)]F^{(1)}(\eta)$ as $F^{(2)}(\eta)$.  Then,
Eqs.~\eref{Wronskian} and \eref{F-Wronskian} imply
\begin{equation}\label{AB-relation}
2A^{(T)}(p)B^{(T)}(p) = \kappa^2 \eqend{.}
\end{equation}
In most of important applications, such as slow-roll inflation, the `positive-frequency' solution $f_p(\eta)$  is chosen
such that
\begin{equation}\label{Ap}
A^{(T)}(p)  \approx  \frac{C}{p^\nu}\eqend{,}\ \ \nu > 0\eqend{.}
\end{equation}
Then, by \eref{AB-relation}, we find
$B^{(T)}(p)\sim p^{\nu}$ for small \(p\).

We note that there is some freedom in distributing the $p$-dependence between
$A^{(T)}(p)$ and $F^{(1)}_p(\eta)$ and between $B^{(T)}(p)$ and
$F^{(2)}_p(\eta)$.  We allow this freedom because in many cases there are standard functions
to be chosen as $F^{(1)}_p(\eta)$ and $F^{(2)}_p(\eta)$.
For example, if $a(\eta)=1$ (flat space), then
we can choose $F^{(1)}_p(\eta) = \cos p\eta$, $F^{(2)}_p(\eta) = p^{-1} \sin p\eta$ (with $\eta \in \mathbb{R}$) and
\begin{equation}
f^{({\scriptstyle {\rm flat}})}_p(\eta) = \frac{\mathrm{i}\kappa}{\sqrt{p}}F^{(1)}_p(\eta) + \kappa \sqrt{p}F^{(2)}_p(\eta)\eqend{,}
\end{equation}
so that $C = \kappa$ and $\nu = 1/2$.

Notice that if $\nu \geqslant (n-1)/2$, then the two-point correlation function
$\Delta_{iji'j'}(\eta,\mathbf{x};\eta',\mathbf{x}')$ is
IR divergent because then the integrand in (\ref{two-point}) will behave like $p^{-2\nu}$, $2\nu \geqslant n-1$.
In the next section we show that large gauge transformations can be used to
make the integrand less singular in the small $p$ limit so that in many applications
the IR divergences can be eliminated by large gauge transformations.


\section{The gauge transformations for the tensor perturbations}\label{gauge-transformation}

The linear gauge transformation for $h_{\mu\nu} = a^2 H_{\mu\nu}$,
\begin{equation}
 \delta_{\tilde{\xi}} h_{\mu\nu} = \tilde{\xi}^\alpha \partial_\alpha g_{\mu\nu} +
(\partial_\mu \tilde{\xi}^\alpha) g_{\alpha\nu} + (\partial_\nu \tilde{\xi}^\alpha) g_{\mu\alpha}\eqend{,}  \label{mainGT}
\end{equation}
can be given, with the definition $\tilde{\xi}_\alpha =a^2(\eta)\xi_\alpha$, as
\begin{equation}
\delta_\xi H_{\mu\nu} = \partial_\mu \xi_\nu + \partial_\nu \xi_\mu - 2 H a \eta_{\mu\nu} \xi_0\eqend{.}
\end{equation}
We show in this section that one can choose $\xi_\alpha$ for each mode function $\gamma_{ij}^{(s,\mathbf{p})}(\eta,\mathbf{x})$ such that the
integrand for the two-point function $\Delta_{iji'j'}(\eta,\mathbf{x};\eta', \mathbf{x}')$ has the power of $p$ reduced by $4$ for small $p$.  That is,
if $(n-1)/2 \leqslant \nu < (n+3)/2$,
then, although the graviton two-point function is IR divergent with the mode functions
$\gamma_{ij}^{(s,\mathbf{p})}(\eta,\mathbf{x})$ behaving like $p^{-\nu}$ for small $p$,
it will be IR finite with the gauge-transformed mode functions.
The gauge transformation we use is given by $\xi_0 = 0$ and
\begin{eqnarray}\label{GT}
\xi_i = -  \frac{\mathrm{i}}{2}A^{(T)}(p) F^{(1)}_{0} (\eta) \left[ \epsilon_{il}^{(s)} (\bp) x^l (1 + \mathrm{i} \bp
\cdot \textbf{x} ) -
\frac{\mathrm{i}}{2}  \epsilon_{lm}^{(s)} (\bp)  x^l x^m p_i  \right]
\mathrm{e}^{- \rho^2 p^2} \eqend{,} \nonumber \\
\end{eqnarray}
where the factor $\mathrm{e}^{-\rho^2 p^2}$, with $\rho$ a positive constant, has been inserted
in order not to introduce
spurious ultraviolet divergences. The polarisation tensors $\epsilon_{ij}^{(s)}(\bp)$ have been defined before
[see the sentence containing (\ref{polarisation-tensor-normalise})].  Notice that they depend only on the direction of
\( \bp \) and not on its magnitude. The part of order \( p^0 \) inside the square brackets represents the
global shear transformation used in~\cite{IRdivpaper}.
The part of order $p$ was obtained by determining the coefficients $\alpha$ and $\beta$ in the general ansatz
\( \alpha\epsilon_{il}^{(s)}(\mathbf{p})x^l \mathbf{p}\cdot\mathbf{x} + \beta\epsilon_{lm}^{(s)}(\mathbf{p})x^lx^m p_i \), which is linear in \( \epsilon^{(s)}_{ij}(\mathbf{p})\) and \( p_i \) and quadratic in \( x^i \).
This is a large gauge transformation in the sense that $\xi_i$ does not tend to zero as $|\mathbf{x}|\to\infty$. In fact it
diverges in this limit.

Now,
\begin{eqnarray}
\delta \gamma_{ij}^{(s,\mathbf{p})}(\eta,\mathbf{x})
& = & \partial_i \xi_j + \partial_j \xi_i \nonumber \\
& = & - \mathrm{i}A^{(T)}(p)\epsilon_{ij}^{(s)}(\mathbf{p})(1+\mathrm{i}\mathbf{p}\cdot\mathbf{x})\mathrm{e}^{-\rho^2p^2}\eqend{,}
\end{eqnarray}
where we used $F^{(1)}_0(\eta) = 1$. Notice that $\delta\gamma_{ij}^{(s,\mathbf{p})}$ is transverse and traceless.
Thus, this gauge transformation does not violate the gauge conditions we have imposed, though the transformed field
no longer has a non-singular Fourier transform.  Now, we find  the transformed mode functions as
\begin{eqnarray}
\tilde{\gamma}_{ij}^{(s,\mathbf{p})}(\eta,\mathbf{x}) &= & \epsilon^{(s)}_{ij}(\mathbf{p})f_p(\mathbf{p})\mathrm{e}^{\mathrm{i}\mathbf{p}\cdot\mathbf{x}} - \mathrm{i}A^{(T)}(p) \epsilon_{ij}^{(s)}(\mathbf{p})(1+\mathrm{i}\mathbf{p}\cdot\mathbf{x}) \mathrm{e}^{-\rho^2 p^2} \nonumber \\
& = & \mathrm{i}\epsilon_{ij}^{(s)} (\mathbf{p}) \Big\{ A^{(T)}(p) \left[ F_p^{(1)}(\eta) - F_0^{(1)}(\eta) \right]
\left( 1 + \mathrm{i}\mathbf{p}\cdot\mathbf{x}\right) \nonumber \\
&& \ \ \ \ \ \ \ \ \ \ \ + A^{(T)}(p)F^{(1)}_p(\eta)\left( \mathrm{e}^{\mathrm{i}\mathbf{p}\cdot\mathbf{x}} - 1 - \mathrm{i}\mathbf{p}\cdot\mathbf{x}\right) \nonumber\\
&& \ \ \ \ \ \ \ \ \ \ \ - A^{(T)}(p)(1+\mathrm{i}\mathbf{p}\cdot\mathbf{x})(\mathrm{e}^{-\rho^2 p^2} - 1)
\nonumber\\
&& \ \ \ \ \ \ \ \ \ \ \ + B^{(T)}(p)F^{(2)}_p(\eta) \mathrm{e}^{\mathrm{i}\mathbf{p}\cdot\mathbf{x}} \Big\} \nonumber \\
& = & \epsilon_{ij}^{(s)}(\mathbf{p})\left[ A^{(T)}(p)\times \mathcal{O}(p^2) + B^{(T)}(p)F_p^{(2)}(\eta)
\mathrm{e}^{\mathrm{i}\mathbf{p}\cdot\mathbf{x}}\right]\eqend{,} \label{gamma-tilde}
\end{eqnarray}
where we have used (\ref{FI}).
Now, suppose $A^{(T)}(p) \approx C/p^\nu$, $\nu \geqslant (n-1)/2$, 
in the small $p$ limit so that the graviton two-point function is IR divergent. 
As a result, due to (\ref{AB-relation}),
$B^{(T)}(p) \sim p^\nu$ for small \(p\).  Then,
the original mode function $\gamma_{ij}^{(s,\mathbf{p})}(\eta,\mathbf{x})$ behaves like $p^{-\nu}$ whereas the transformed mode function
$\tilde{\gamma}_{ij}^{(s,\mathbf{p})}$ behaves like $p^{-\nu+2}$.  Thus, if
\begin{equation}
\frac{n-1}{2} \leqslant \nu < \frac{n+3}{2}\eqend{,} \label{sigma-condition}
\end{equation}
then the original
graviton two-point function $\Delta_{iji'j'}(\eta,\mathbf{x};\eta',\mathbf{x}')$
given by (\ref{two-point}) is IR divergent whereas the
transformed one with $\gamma_{ij}^{(s,\mathbf{p})}(\eta,\mathbf{x})$ replaced by
$\tilde{\gamma}_{ij}^{(s,\mathbf{p})}(\eta,\mathbf{x})$ is
IR finite.  Thus, for this range of $\nu$, the IR divergences of the graviton two-point function can be removed by
large gauge transformations.

Below, we apply the general observation described above to the particular case with
 $a(\eta) = (-\eta/\eta_0)^{-\lambda}$, where $\eta_0$ and
$\lambda$ are positive constants, and where $\eta$ runs from $-\infty$ to $0$.  In this case the field equation
\eref{eq-mo} becomes
\begin{equation}
f_p''(\eta) + \frac{(n-2)\lambda}{\eta}f_p'(\eta) + p^2f_p(\eta) = 0\eqend{.} \label{eq-mo-const-epsilon}
\end{equation}
A solution to this equation is
\begin{eqnarray}
f_p(\eta) = C^{(T)}(p)(-p\eta)^{\nu} H_\nu^{(1)}(-p\eta)\eqend{,} \label{def-f} \\
\nu  = \frac{1}{2}\left[ 1 + (n-2)\lambda\right]\eqend{,}
\end{eqnarray}
where $H_\nu^{(1)}(z)$ is the Hankel function of the first kind.
The constant $\lambda$ can be related to the slow-roll parameter $\epsilon$ defined by (\ref{def-epsilon}) as
\begin{equation}
\epsilon =  1- \frac{1}{\lambda}\eqend{.}
\end{equation}
We note that both the radiation phase and matter phase correspond to fixed values of \(\epsilon\)~\cite{sasaki}.
Since $\epsilon$ is time independent, the slow-roll parameter 
\begin{equation}
\delta: = \frac{\epsilon'}{2Ha\epsilon} = 0\eqend{.} \label{def-delta}
\end{equation}
The mode functions
$\gamma_{ij}^{(s,\mathbf{p})}(\eta,\mathbf{x})$ defined by \eref{tensor-mode}
transform under the scaling $(\eta,\mathbf{x}) \to (\alpha\eta,\alpha\mathbf{x})$, where $\alpha$
is a positive constant, as
\begin{equation}
\gamma_{ij}^{(s,\mathbf{p})}(\alpha\eta,\alpha\mathbf{x})
= f(\alpha) \gamma_{ij}^{(s,\alpha\mathbf{p})}(\eta,\mathbf{x})\eqend{,}
\end{equation}
where 
\begin{equation}
f(\alpha) = \frac{C^T(p)}{C^T(\alpha p)} \eqend{.}
\end{equation}
Thus, if one defines the vacuum state $|0\rangle$ in section \ref{quantisation} by adopting the function $f_p(\eta)$ defined by (\ref{def-f}), then it
respects the scaling symmetry $(\eta,\mathbf{x}) \to (\alpha\eta,\alpha\mathbf{x})$.   This state is the natural vacuum state in this sense and is generally adopted 
for the slow-roll inflationary models, for example.
(The derivatives of the graviton two-point function that are IR finite acquire a constant factor under this
scaling for slow-roll inflation~\cite{InflationGTpre}.)
For this reason we study this case below.

The normalisation constant $C^{(T)}(p)$ can readily be found up to an overall phase factor
by using the large $\eta$ limit of (\ref{Wronskian}) with
\begin{equation}
H_\nu^{(1)}(z) \approx \sqrt{\frac{2}{\pi z}}\mathrm{e}^{\mathrm{i}(z-\frac{\pi}{2}\nu - \frac{\pi}{4})}\eqend{.}
\end{equation}
We find
\begin{equation}
C^{(T)}(p)=  - \kappa \frac{\sqrt{\pi\eta_0}}{\sqrt{2}(p\eta_0)^\nu}\eqend{.}  \label{def-Cp}
\end{equation}
Thus, $C^{(T)}(p)\sim p^{-\nu} =  p^{-\left[1+(n-2)\lambda\right]/2}$ for small $p$.  We can write $f_p(\eta)$ in the form (\ref{convention}) with
\begin{eqnarray}
F^{(1)}_p(\eta) & = & - \frac{\pi}{2^\nu\Gamma(\nu)}(-p\eta)^\nu Y_\nu(-p\eta)\eqend{,} \label{F1_p}\\
F^{(2)}_p(\eta) & = & - \frac{2^{\nu -1} \Gamma(\nu)\eta_0}{(p\eta_0)^{2\nu}}(-p\eta)^\nu J_\nu(-p\eta)\eqend{,} \label{F2_p}\\
A^{(T)}(p) & =  & \kappa \sqrt{\frac{\eta_0}{\pi}}\frac{2^{\nu-\frac{1}{2}}\Gamma(\nu)}{(p\eta_0)^\nu}\eqend{,} \label{AT_p}\\
B^{(T)}(p) & = & \kappa \sqrt{\frac{\pi}{\eta_0}}\frac{(p\eta_0)^\nu}{2^{\nu-\frac{1}{2}} \Gamma(\nu)}\eqend{,} \label{BT_p}
\end{eqnarray}
where $J_\nu(z)$ and $Y_\nu(z)$ are the Bessel functions of the first and second kinds, respectively.

From (\ref{sigma-condition}) we find that, if
\begin{equation}
1 \leqslant \lambda < \frac{n+2}{n-2}\eqend{,} \label{lambda-condition}
\end{equation}
then the two-point function (\ref{two-point}) is IR divergent, but that the one after
large gauge transformations given by~\eref{GT} is IR finite.  In terms of the slow-roll
parameter \(\epsilon\), this condition can be written as
\begin{equation}
0 \leqslant \epsilon < \frac{4}{n+2}\eqend{,}  \label{positive-lambda}
\end{equation}
the case $\epsilon=0$ being the de~Sitter limit.   Interestingly, the spacetime with $\epsilon < 0$ ($0 < \lambda <1$)  (Big Rip spacetime~\cite{BigRip})
causes no IR problems.

Let us briefly discuss the case with $\lambda < 0$.  (The case $\lambda = 0$ gives Minkowski space.)
In this case we have
$a(\eta) = (\eta/\eta_0)^{|\lambda|}$ and the variable $\eta$ runs from $0$ to $+\infty$ in order to have an expanding universe.
The function $f_p(\eta)$ that  we can adopt in this case is
\begin{equation}
f_p^{(\lambda<0)}(\eta) = -C^{(T)}(p)(p\eta)^\nu H^{(2)}_\nu(p\eta)\eqend{,}
\end{equation}
where $C^{(T)}(p)$ is given by (\ref{def-Cp}) and where $\nu = \frac{1}{2}\left| (n-2)\lambda + 1\right|$.
In a way similar to the positive $\lambda$ case, we
find that the two-point function \eref{two-point} is IR divergent but can be rendered IR finite by the large gauge transformations given by \eref{GT} if
$2- 6/(n+4) < \epsilon \leqslant 2- 2/n$.  Combining this case and \eref{positive-lambda} for positive $\lambda$, we find that
the two-point function \eref{two-point} is IR divergent if $0 \leqslant \epsilon \leqslant 2 - 2/n$, but the gauge transformations \eref{GT} render it IR finite unless
$4/(n+2) \leqslant \epsilon\leqslant 2 - 6/(n+4)$.

It is known that the two-point function for the linearised Weyl tensor, which is a local gauge invariant, is also IR divergent if and only if
$4/(n+2) \leqslant \epsilon \leqslant 2 - 6/(n+4)$~\cite{bound} for the vacuum state chosen here.  This implies that it is impossible to render the graviton
two-point function IR finite by any gauge transformations outside this range of values for $\epsilon$ because the linearised Weyl tensor is invariant under
any gauge transformation, large or otherwise.

For $0 \leqslant \epsilon \ll 1$, i.e.\ for slow-roll inflationary FLRW universe, our result clearly shows that
the IR divergence of the two-point function for the tensor perturbations
can be eliminated by large gauge transformations.  In the next section we show that we can do the same for the scalar
perturbations in this spacetime.

\section{Scalar perturbations in single-field inflation}\label{single-field}

We now consider inflationary FLRW spacetime such that the inflation is driven by a scalar (inflaton) field
$\tilde{\phi}(\eta,\mathbf{x}) = \phi(\eta)+\psi(\eta,\mathbf{x})$ with
the background $\phi(\eta)$ depending only on conformal time $\eta$.  The linear gauge transformations are given by \eref{mainGT} on the gravitational field and
\begin{eqnarray}
\delta \psi = \mathcal{L}_{\tilde{\xi}} \phi = \tilde{\xi}^\mu \partial_\mu \phi \eqend{,}
\end{eqnarray}
on the perturbation $\psi$ of the inflaton. We start by considering the parts of the components of the graviton
$h_{\mu\nu}$ and
inflaton $\psi$ invariant under \emph{local} gauge transformations
and then write down the action in terms of those gauge invariant variables. It is only after writing down
the action that we make a gauge choice.

The components of the gravitational field $h_{\mu\nu}$
can be written in terms of 4 gauge invariant quantities. There is one gauge invariant tensor, \(H_{kl}\), which is the transverse traceless part of \(h_{kl}\).
Additionally, we have one gauge invariant transverse vector, denoted \(V_k\). We have two gauge invariant scalars, labelled \(S\) and \(\Sigma\); we shall see later
that there is only one dynamical gauge invariant scalar, which is a linear combination of \(S\) and \(\Sigma\). (For full details of this decomposition, see \cite{Weyl}.)
Writing the components of the perturbation $h_{\mu\nu}$ in terms of these quantities gives
\begin{eqnarray}
h_{00} &= S + 2 X'_0 + 2 H a X_0 \eqend{,} \label{zerozeroinv} \\
h_{0k} &= V_k + X'_k + \partial_k X_0 \eqend{,} \label{zerokinv} \\
h_{kl} &= H_{kl} + \delta_{kl} \Sigma + 2 \partial_{(k} X_{l)} - 2 H a \delta_{kl} X_0 \label{klinv} \eqend{.}
\end{eqnarray}
In this form the gauge transformation~\eref{mainGT} can be attributed to that of the fields $X_\mu$:
\begin{eqnarray}\label{gauge-trans-X}
\delta X_\mu = \xi_\mu \eqend{.}
\end{eqnarray}
We similarly write the perturbation $\psi$ of the inflaton in terms of this vector \( X_\mu\) and another gauge invariant scalar \(\Psi\) as
\begin{eqnarray}
\psi = \Psi - X_0 \phi'  \label{psiinv} \eqend{.}
\end{eqnarray}
In fact, this equation defines the scalar \(\Psi\).

Now, we consider the Einstein-Hilbert action, along with the action for a minimally coupled scalar field \( \tilde{\phi}\),
\begin{eqnarray}
I = \frac{1}{\kappa ^2} \int \tilde{R} \sqrt{-\tilde{g}} \mathrm{d}^n x - \frac{1}{2} \int \sqrt{-\tilde{g}} \mathrm{d}^n x \left[ \tilde{g}^{\mu \nu} (\partial_{\mu} \tilde{\phi} )( \partial_{\nu} \tilde{\phi} ) + V(\tilde{\phi} )\right] \eqend{,}
\end{eqnarray}
for some potential \(V(\phi)\).   One expands this action to second order and substitutes \eref{zerozeroinv}-\eref{klinv} and \eref{psiinv}
into the resulting quadratic action.
Varying the action with respect to \(V_k\) and \(S\)
results in the following
constraint equations~\cite{Weyl}:
\begin{eqnarray}
V_k &= 0 \eqend{,} \label{VEoM} \\
S &= (n-3) \Sigma \eqend{.} \label{SEoM}
\end{eqnarray}
Here we are working in the space of functions where the Laplacian $\laplace$ is invertible.
Then, by  introducing the Sasaki-Mukhanov variable~\cite{sasaki},
\begin{eqnarray} \label{sasakimukhanov}
Q \equiv \frac{2Ha}{\phi'} \Psi - \Sigma \eqend{,}
\end{eqnarray}
 which describes the scalar perturbations, and the non-dynamical variables \(S\) and \(V_k\) through
Eqs.~\eref{VEoM} and \eref{SEoM}, one finds the following action~\cite{Weyl}:
\begin{eqnarray} \label{actionexpand}
I^{(2)} =& \frac{1}{4 \kappa^2} \int \left[ H'_{kl}H'_{kl} + H_{kl} \triangle H_{kl} \right] a^{n-2} \mathrm{d}^n x  \nonumber \\
& + \frac{n-2}{4\kappa^2} \int \left[ Q^{\prime 2} + Q \triangle Q \right] \epsilon a^{n-2} \mathrm{d}^n x\nonumber \\
&+ \frac{n-2}{4\kappa^2} \int \frac{ \left( \triangle \Sigma - \frac{H'}{H} Q' \right)^2}{(n-1 - \epsilon)H^2a^2} a^{n-2} \mathrm{d}^n x \eqend{.}
\end{eqnarray}
Lagrange's equation for $\Sigma$ is a constraint equation:
\begin{equation}
 \laplace \Sigma = - \epsilon H a Q'\eqend{.} \label{SigmaEoM}
\end{equation}
After this constraint is imposed, the field equation for $Q$ takes the following form:
\begin{equation} \label{Q-fieldeq}
Q'' + \left( n-2 +2 \delta  \right) Ha Q' - \laplace Q = 0\eqend{,}
\end{equation}
where the slow-roll parameter \(\delta\) is defined by~\eref{def-delta}.

The tensor perturbation $H_{kl}$ here can be treated in exactly the same way as in sections~\ref{tensor-perturbation} and \ref{quantisation} and the results
obtained there will apply for single-field inflation as well.  The  Sasaki-Mukhanov variable $Q$ can be quantised in the standard manner.  One finds
\begin{eqnarray}
Q (\eta, \bx) &= \int \frac{\mathrm{d}^{n-1}\mathbf{p}}{(2\pi)^{n-1}} a(\mathbf{p}) q_p(\eta) e^{i\mathbf{p}\cdot\mathbf{x}} + h.c. \eqend{,}
\end{eqnarray}
where the function $q_p(\eta)$ satisfies
\begin{equation}
q''_p(\eta) + \left( n-2 +2 \delta  \right)Ha q'_p(\eta) + p^2 q_p(\eta) = 0\eqend{.} \label{qp-eq}
\end{equation}
By considering the Wronskian for this equation, we find that we can require
\begin{equation}
q_p^*(\eta) q_p^{\prime} (\eta) - q_p(\eta)q_p^{*\prime}(\eta)
= \frac{2\mathrm{i}\kappa^2}{(n-2)\epsilon(\eta) a^{n-2}(\eta)}\eqend{.}  \label{Wronskian-scalar}
\end{equation}
The operators $a(\mathbf{p})$ and $a^\dagger(\mathbf{p})$ then satisfy
\begin{eqnarray}
[a (\bp), a^\dagger (\bp \p) ] = (2\pi)^{n-1} \delta^{n-1} ( \bp - \bp \p) \eqend{.}
\end{eqnarray}
Defining the vacuum state $|0\rangle$ by requiring that $a(\mathbf{p})|0\rangle = 0$ for all $\mathbf{p}$, we find the two-point function for $Q(\eta,\mathbf{x})$
as
\begin{eqnarray}
\Delta(\eta,\mathbf{x};\eta',\mathbf{x}')
& : = & \langle 0| Q(\eta,\mathbf{x})Q(\eta',\mathbf{x}')|0\rangle \nonumber \\
& = & \int \frac{\mathrm{d}^{n-1}\mathbf{p}}{(2\pi)^{n-1}} q_p(\eta)q^*_p(\eta') \mathrm{e}^{\mathrm{i}\mathbf{p}\cdot(\mathbf{x}-\mathbf{x}')}
\eqend{.} \label{Q-2p-function}
\end{eqnarray}

Now, we analyse the solutions $q_p(\eta)$ for small $p$.  For $p=0$, two independent real solutions $q_0(\eta) =
Q_0^{(1)}(\eta)$ and $Q_0^{(2)}(\eta)$
can be chosen as
\begin{eqnarray}
Q_0^{(1)}(\eta) &=  & 1,\\
Q_0^{(2)}(\eta) & = &  \int \frac{\mathrm{d}\eta}{\epsilon(\eta)a^{n-1}(\eta)}\eqend{,}  \label{Q0-2}
\end{eqnarray}
where the constant of integration in \eref{Q0-2} is suitably chosen.  As in the tensor case,
two independent solutions $Q^{(1)}_p(\eta)$ and $Q^{(2)}_p(\eta)$ can be chosen for
nonzero $p$ such that
\begin{equation}
Q_p^{(1)}(\eta) \frac{\mathrm{d}Q_p^{(2)}(\eta)}{\mathrm{d}\eta} - \frac{\mathrm{d}Q_p^{(1)}(\eta)}{\mathrm{d}\eta}Q_p^{(2)}(\eta)
= \frac{1}{\epsilon(\eta)a^{n-2}(\eta)}\eqend{,}
\end{equation}
and that
\begin{equation}
Q_p^{(I)}(\eta) = Q_0^{(I)}(\eta) + \mathcal{O}(p^2),\ \ I=1,2\eqend{.}
\end{equation}
Again, in most applications, such as slow-roll inflation, the solutions $q_p(\eta)$ are chosen as
\begin{equation}
q_p(\eta) = \mathrm{i}A^{(S)}(p)Q_p^{(1)}(\eta) + B^{(S)}(p)Q_p^{(2)}(\eta)\eqend{,}
\end{equation}
such that
\begin{equation}
A^{(S)}(p) \approx \frac{C'}{p^{\nu'}}\eqend{,}
\end{equation}
where $C'$ and $\nu'$ are positive constants, for small $p$.  As in the tensor case, we can let
$B^{(S)}(p)\sim p^{\nu'}$ as $p\to 0$.
If $\nu' \geqslant (n-1)/2$, then the two-point
function $\Delta(\eta,\mathbf{x};\eta',\mathbf{x}')$ given by \eref{Q-2p-function} is IR divergent.  The IR divergences of the two-point functions for
the tensor perturbations and Sasaki-Mukhanov variable are reflected in those for the graviton $h_{\mu\nu}$ and the inflaton $\psi$.  In the next section we shall
see that these IR divergences can be gauged away by large gauge transformations if $\nu' < (n+3)/2$.


\section{IR divergences in single-field inflation}\label{IR-in-single-field}

In this section we show that, even if the two-point functions for the tensor and scalar perturbations are IR divergent,
one can eliminate these divergences by large
gauge transformations as long as they are not very severe.
Since the mechanism for the IR-divergence elimination has been discussed already for the tensor
perturbations in section~\ref{gauge-transformation}, here we focus on the scalar perturbations.

We start with the graviton field in the gauge where the perturbation $\psi$ in the scalar field is set to $0$.  The gauge
is fixed by choosing the fields $X_\mu$ because of~\eref{gauge-trans-X}.  We choose them as follows:
%
\begin{eqnarray}
X_0 &= \frac{\Psi}{\phi'}\eqend{,} \label{scalarvanishgaugetime} \\
X_k &= 0 \eqend{.} \label{scalarvanishgaugespace}
\end{eqnarray}
Then we find,  from \eref{zerozeroinv}-\eref{klinv}, that
\begin{eqnarray}
h_{00} &= \frac{1}{H a} Q' \eqend{,} \label{hzerozero}\\
h_{0k} &= \frac{1}{2} \partial_k \left( \frac{1}{H a} Q - \epsilon \laplace^{-1} Q'\right) \eqend{,} \\
h_{kl} &= H_{kl} - \delta_{kl} Q \eqend{,} \label{hkl}\\
\psi &= 0 \eqend{.} \label{psi_gf}
\end{eqnarray}
The derivation of these formulas is given in \ref{Appendix-A}.
This gauge corresponds to imposing the following conditions on the components of the graviton and scalar fields:
\begin{eqnarray}
\partial^l \left( h_{kl} - \frac{1}{n-1} \delta_{kl} \delta^{ij} h_{ij} \right) = 0 \eqend{,} \label{gcone} \\
\psi = 0 \eqend{.} \label{gctwo}
\end{eqnarray}
Note that the condition \eref{gcone} states that the traceless part of $h_{kl}$ is transverse.  Thus, the
field components $h_{\mu\nu}$ can be expressed as
\begin{eqnarray}
h_{00}(\eta,\mathbf{x})
& = & \int \frac{\mathrm{d}^{n-1}\mathbf{p}}{(2\pi)^{n-1}}
a(\mathbf{p}) \gamma_{00}^{(\mathbf{p})} (\eta,\mathbf{x}) + \textrm{h.c.}\eqend{,}\\
h_{0k}(\eta,\mathbf{x})
& = & \int \frac{\mathrm{d}^{n-1}\mathbf{p}}{(2\pi)^{n-1}}
a(\mathbf{p}) \gamma_{0k}^{(\mathbf{p})} (\eta,\mathbf{x})  + \textrm{h.c.}\eqend{,}\\
h_{kl}(\eta,\mathbf{x})
& = &  \int \frac{\mathrm{d}^{n-1}\mathbf{p}}{(2\pi)^{n-1}}
\left[ a(\mathbf{p}) \gamma^{(\mathbf{p})}_{kl}(\eta,\mathbf{x})+
\sum_{s} a_s(\mathbf{p})\gamma_{kl}^{(s,\mathbf{p})}(\eta,\mathbf{x}) \right]
\mathrm{e}^{\mathrm{i}\mathbf{p}\cdot\mathbf{x}} + \textrm{h.c.}\eqend{,}\nonumber \\
\end{eqnarray}
where $\gamma^{(s,\mathbf{p})}(\eta,\mathbf{x})$ are defined by \eref{tensor-mode}, and where
\begin{eqnarray}
\gamma_{00}^{(\mathbf{p})}(\eta,\mathbf{x}) & = & \frac{1}{H(\eta)a(\eta)}q'_p(\eta)\mathrm{e}^{\mathrm{i}\mathbf{p}\cdot\mathbf{x}}\eqend{,} \label{g00} \\
\gamma_{0k}^{(\mathbf{p})}(\eta,\mathbf{x}) & = & \frac{\mathrm{i}}{2} p_k \left[ q_p(\eta)  + \frac{\epsilon(\eta)}{p^2}q'_p(\eta)\right]
\mathrm{e}^{\mathrm{i}\mathbf{p}\cdot\mathbf{x}}
\eqend{,} \label{g0k} \\
\gamma_{kl}^{(\mathbf{p})}(\eta,\mathbf{x}) & = & - \delta_{kl}q_p(\eta)\mathrm{e}^{\mathrm{i}\mathbf{p}\cdot\mathbf{x}}\eqend{.} \label{gkl}
\end{eqnarray}

The space components of the two-point function of $h_{\mu\nu}$ are
\begin{eqnarray}
& \langle 0|h_{kl}(\eta,\mathbf{x})h_{k'l'}(\eta',\mathbf{x}')|0\rangle \nonumber \\
& = \int
\frac{\mathrm{d}^{n-1}\mathbf{p}}{(2\pi)^{n-1}}
\left[ \gamma_{kl}^{(\mathbf{p})}(\eta,\mathbf{x}) \gamma_{k'l'}^{(\mathbf{p})*}(\eta',\mathbf{x}') +
\sum_{s}\gamma_{kl}^{(s,\mathbf{p})}(\eta,\mathbf{x})
\gamma_{k'l'}^{(s,\mathbf{p})*}(\eta',\mathbf{x}')
\right]\eqend{.}   \label{2pt-space}  
\end{eqnarray}
The other components are given by
\begin{equation}
\langle 0|h_{\mu\nu}(\eta,\mathbf{x})h_{\mu'\nu'}(\eta',\mathbf{x}')|0\rangle
= \int
\frac{\mathrm{d}^{n-1}\mathbf{p}}{(2\pi)^{n-1}}
\gamma_{\mu\nu}^{(\mathbf{p})}(\eta,\mathbf{x}) \gamma_{\mu'\nu'}^{(\mathbf{p})}(\eta',\mathbf{x}')
\eqend{,} \label{2pt-other}
\end{equation}
where at least one of the indices $\mu$, $\nu$, $\mu'$ and $\nu'$ is the time index $0$.
The IR properties of these two-point functions are determined by the
small-\(p\) behaviour of the integrand for the \(p\)-integral.

Assuming the properties of $q_p(\eta)$ stated in section~\ref{single-field}, in particular
that for small \(p\) one has
$q_p(\eta) \approx \mathrm{i}A^{(S)}(p)\left[ 1 + \mathcal{O}(p^2)\right]$, with
$A^{(S)}(p) \sim 1/p^{\nu'}$, where $\nu' \geqslant (n-1)/2$,
and that $B^{(S)}(p)\sim p^{\nu'}$, it is clear that the small-\(p\) behaviour of the
derivative, $q_p'(\eta)$, is better by a factor of $p^2$,
i.e.\ $q_p'(\eta) \sim 1/p^{\nu'-2}$ for small $p$.  As a result, we find
$\gamma^{(\mathbf{p})}_{00}(\eta,\mathbf{x}) \sim 1/p^{\nu'-2}$ and
$\gamma^{(\mathbf{p})}_{0k}(\eta,\mathbf{x})\sim 1/p^{\nu'-1}$ for small $p$.  We now show
that large gauge transformations similar to those given by \eref{GT} can be used so that all functions
$\gamma^{(\mathbf{p})}_{\mu\nu}(\eta,\mathbf{x})$ and
$\gamma_{\mu \nu}^{(s,\mathbf{p})}(\eta,\mathbf{x})$
are modified to behave like $1/p^{\nu'-2}$ rather than $1/p^{\nu'}$. 

Let us define
\begin{equation}
\tilde{Q}_p^{(1)}(\eta) : = \frac{1}{p^2}Q_p^{(1)\prime}(\eta)\eqend{,}
\end{equation}
and we note that the function $\tilde{Q}_0^{(1)}(\eta) : = \lim_{p\to 0}\tilde{Q}_p^{(1)}(\eta)$ is well defined
because $Q_p^{(1)\prime}(\eta) = \mathcal{O}(p^2)$.
Then,
\begin{eqnarray}
\gamma_{00}^{(\mathbf{p})}(\eta,\mathbf{x}) & \approx & \frac{\mathrm{i}}{H(\eta)a(\eta)}p^2 A^{(S)}(p)\tilde{Q}_p^{(1)}(\eta)
\mathrm{e}^{\mathrm{i}\mathbf{p}\cdot\mathbf{x}}\eqend{,} \\
\gamma_{0k}^{(\mathbf{p})}(\eta,\mathbf{x}) & \approx & - \frac{1}{2} p_k A^{(S)}(p)\left[ Q_p^{(1)}(\eta) + \epsilon(\eta)\tilde{Q}_p^{(1)}(\eta)\right]
\mathrm{e}^{\mathrm{i}\mathbf{p}\cdot\mathbf{x}}\eqend{,}\\
\gamma_{kl}^{(\mathbf{p})}(\eta,\mathbf{x}) & \approx & - \mathrm{i}\delta_{kl}A^{(S)}(p)Q_p^{(1)}(\eta)
\mathrm{e}^{\mathrm{i}\mathbf{p}\cdot\mathbf{x}}\eqend{,}
\end{eqnarray}
for small $p$. The tensor modes are modified as described in section \ref{gauge-transformation}. For the scalar modes, we make the large gauge transformation with $\xi_0 = 0$ and
\begin{eqnarray}
\xi_i =& \frac{\mathrm{i}}{2}A^{(S)}(p) Q_0^{(1)} (\eta) \left[ (1+\mathrm{i} \bp \cdot \bx )x_i 
- \frac{\mathrm{i}}{2} p_i x^2 \right] e^{- \rho^2 p^2} \nonumber \\
& + \frac{1}{2} A^{(S)}(p) p_i \int \mathrm{d} \eta \left[ Q_0^{(1)}(\eta) + \epsilon(\eta) \tilde{Q}_0^{(1)}(\eta)\right] e^{- \rho^2 p^2} \eqend{.}
\label{inflation-gauge-transformation}
\end{eqnarray}
The 
line was obtained
in a way similar to the method
used in section \ref{gauge-transformation}.
The 
second line gauges away the \(\mathcal{O}(p)\) term in 
\(\gamma_{0k}^{(\mathbf{p})}(\eta,\mathbf{x})\).
Then the two-point functions \eref{2pt-space} and \eref{2pt-other} are modified in such a way that the tensor mode functions
$\gamma_{kl}^{(s,\mathbf{p})}(\eta,\mathbf{x})$ are replaced by $\tilde{\gamma}_{kl}^{(s,\mathbf{p})}(\eta,\mathbf{x})$
given by \eref{gamma-tilde} and that the functions $\gamma_{\mu\nu}^{(\mathbf{p})}(\eta,\mathbf{x})$, given by \eref{g00} - \eref{gkl}, are replaced by
\begin{eqnarray}
\tilde{\gamma}_{00}^{(\mathbf{p})}(\eta,\mathbf{x}) & = & \gamma_{00}^{(\mathbf{p})}(\eta,\mathbf{x})\eqend{,}\\
\tilde{\gamma}_{0k}^{(\mathbf{p})}(\eta,\mathbf{x}) & = & - \frac{1}{2} p_k \Big\{
A^{(S)}(p)\left[ Q_p^{(1)}(\eta) - Q_0^{(1)}(\eta)\right] - \mathrm{i}B^{(S)}(p)Q_p^{(2)}(\eta) \nonumber \\
&& \ \ \ \ \ \ \ \ \ \ + \epsilon(\eta)A^{(S)}(p)\left[ \tilde{Q}_p^{(1)}(\eta) - \tilde{Q}_0^{(1)}(\eta)\right] \nonumber \\
&& \ \ \ \ \ \ \ \ \ \ - \mathrm{i} \epsilon(\eta) p^{-2} B^{(S)}(p)Q^{(2)\prime}_p(\eta) \Big\} \mathrm{e}^{\mathrm{i}\mathbf{p}\cdot\mathbf{x}} \nonumber \\
&& - \frac{1}{2} p_kA^{(S)}(p)\left[ Q_0^{(1)}(\eta) + \epsilon(\eta)\tilde{Q}_0^{(1)}(\eta)\right](\mathrm{e}^{\mathrm{i}\mathbf{p}\cdot\mathbf{x}} -
\mathrm{e}^{-\rho^2p^2}) \eqend{,} \\
\tilde{\gamma}_{kl}^{(\mathbf{p})}(\eta,\mathbf{x})& = & -\mathrm{i}\delta_{kl}\Big\{ A^{(S)}(p)\left[ Q_p^{(1)}(\eta)  - Q_0^{(1)}(\eta)\right]
\left( 1 + \mathrm{i}\mathbf{p}\cdot\mathbf{x}\right) \nonumber \\
&& \ \ \ \ \ \ \ \ \ + A^{(S)}(p)Q^{(1)}_p(\eta)\left( \mathrm{e}^{\mathrm{i}\mathbf{p}\cdot\mathbf{x}}
- 1 - \mathrm{i}\mathbf{p}\cdot\mathbf{x}\right) \nonumber\\
&& \ \ \ \ \ \ \ \ \ - A^{(S)}(p)(1+\mathrm{i}\mathbf{p}\cdot\mathbf{x})(\mathrm{e}^{-\rho^2 p^2} - 1)  \nonumber\\
&& \ \ \ \ \ \ \ \ \ - \mathrm{i}B^{(S)}(p)Q^{(2)}_p(\eta) \mathrm{e}^{\mathrm{i}\mathbf{p}\cdot\mathbf{x}} \Big\} \eqend{.}
\end{eqnarray}
Thus,
all components of
$\tilde{\gamma}_{\mu\nu}^{(\mathbf{p})}(\eta,\mathbf{x})$ behave like $p^{\nu' -2}$ or better for  small $p$.  This implies that, although
the two-point function for the metric perturbation is IR divergent if $\textrm{max}(\nu,\nu') \geqslant (n-1)/2$, the two-point function modified
by the large gauge transformations given by \eref{GT} and \eref{inflation-gauge-transformation} is IR divergent only if
$\textrm{max}(\nu,\nu')\geqslant (n+3)/2$.

If the slow-roll parameter $\epsilon\,(>0)$ is constant, i.e.\ if the scale parameter takes the form
$a(\eta) = (-\eta/\eta_0)^{-\lambda}$, then the tensor perturbation $H_{\mu\nu}$ satisfies the same equation as in section~\ref{gauge-transformation},
so the functions $F^{(1)}_p(\eta)$, $F^{(2)}_p(\eta)$, and the constants $A^{(T)}(p)$, $B^{(T)}(p)$, are given by \eref{F1_p}, \eref{F2_p},
\eref{AT_p} and \eref{BT_p}.  As for the scalar perturbation $Q$, the function $q_p(\eta)$ satisfies the same equation as $f_p(\eta)$. Taking into account
the normalisation condition \eref{Wronskian} we find
\begin{equation}
q_p(\eta) = C^{(S)}(p)(-p\eta)^{\nu}H_\nu^{(1)}(-p\eta)\eqend{,}
\end{equation}
where $\nu = [1+(n-2)\lambda]/2$ as before, and
\begin{equation}
C^{(S)}(p)=  - \kappa \frac{\sqrt{\pi\eta_0}}{\sqrt{2(n-2)\epsilon}(p\eta_0)^\nu}\eqend{.}
\end{equation}
Then we can let
\begin{eqnarray}
Q^{(1)}_p(\eta) & = & - \frac{\pi}{2^\nu\Gamma(\nu)}(-p\eta)^\nu Y_\nu(-p\eta)\eqend{,} \label{Q1-p-eta}\\
Q^{(2)}_p(\eta) & = & - \frac{2^{\nu -1} \Gamma(\nu)\eta_0}{(p\eta_0)^{2\nu}}(-p\eta)^\nu J_\nu(-p\eta)\eqend{,} \\
A^{(S)}(p) & =  & \kappa \sqrt{\frac{\eta_0}{\pi(n-2)\epsilon}}\frac{2^{\nu-\frac{1}{2}}\Gamma(\nu)}{(p\eta_0)^\nu}\eqend{,} \\
B^{(S)}(p) & = & \kappa \sqrt{\frac{\pi}{(n-2)\epsilon\eta_0}}\frac{(p\eta_0)^\nu}{2^{\nu-\frac{1}{2}}  \Gamma(\nu)}\eqend{,} \label{BS-p}
\end{eqnarray}
By using the recursion formula for the Bessel functions,
\begin{equation}
\frac{\mathrm{d}\ }{\mathrm{d}z}\left[ z^\nu J_\nu(z)\right] = z^\nu J_{\nu-1}(z)\eqend{,}
\end{equation}
and similarly for $Y_\nu(z)$, we find
\begin{eqnarray}
\tilde{Q}_p^{(1)}(\eta) & = & - \frac{\pi \eta}{2^\nu \Gamma(\nu)}(-p\eta)^{\nu-1}Y_{\nu-1}(-p\eta)\eqend{,} \\
p^{-2}Q_p^{(2)\prime}(\eta) & = & - \frac{2^{\nu-1}\Gamma(\nu)\eta_0\eta}{(p\eta_0)^{2\nu}} (-p\eta)^{\nu-1}
J_{\nu-1}(-p\eta)\eqend{.}
\end{eqnarray}
The function $\tilde{Q}_p^{(1)}(\eta)$ is non-singular as $p\to 0$ as concluded from the general discussion.
The range of the parameter $\epsilon$ for which the IR divergences can be gauged away is the same as that for the
tensor perturbations and given by \eref{lambda-condition} or, equivalently, by \eref{positive-lambda} if $\lambda > 0$.
The discussion for the case $\lambda < 0$ is also exactly the same as the tensor-perturbation case.

The scale factor corresponding to the slow-roll inflation can be written as
\begin{equation}
a(\eta) = (-\eta/\eta_0)^{-\frac{1}{1 - \epsilon} + \epsilon \delta\ln(-\eta/\eta_0)}\eqend{,}  \label{sf-slow-roll}
\end{equation}
where $\epsilon \geqslant 0$, and $\epsilon, |\delta| \ll 1$, 
in a range of $\eta$ where $\epsilon$ and $\delta$ can be treated as constants.\footnote{One cannot write $a(\eta)$ in this form over the long range of $\eta$ for which variations in $\eta$ and $\delta$ need to be taken into account.}
We work to first order in $\epsilon$ and $\delta$.
One readily finds that the slow-roll parameters agree with the parameters $\epsilon$ and $\delta$ in
\eref{sf-slow-roll} to lowest order.  The discussion of the IR divergences for the tensor perturbations
will be exactly the same as in section~\ref{gauge-transformation} except that here
the slow-roll parameter $\epsilon$ is
assumed to be much smaller than $1$.  The discussion for the scalar perturbations is changed slightly if $\delta \neq 0$.
Noting that $Ha = - [(1-\epsilon)\eta]^{-1}$ to first order, one finds that equation~\eref{qp-eq} becomes
\begin{equation}
q_p''(\eta) + \frac{(n-2)(1+\epsilon)+2\delta}{\eta}q_p'(\eta) + p^2 q_p(\eta) = 0\eqend{.}
\end{equation}
By comparing this equation with \eref{eq-mo}, we find that two independent solutions can be chosen as
$q_p(\eta) \propto (-p\eta)^{\nu'} H_{\nu'}^{(1)}(-p\eta)$ and its complex conjugate, where
\begin{equation}
\nu' = \frac{1}{2}[1 + (n-2)(1+\epsilon)+2\delta]\eqend{.}
\end{equation}
The normalisation constant can be found from \eref{Wronskian-scalar}.  By noting that we can write
\begin{equation}
\epsilon(\eta) =\epsilon_0 (-\eta/\eta_0)^{-2\delta}\eqend{,}
\end{equation}
to next leading order in $\epsilon$ and $\delta$, where $\epsilon_0$ is a constant, we find
\begin{equation}
q_p(\eta) = C^{(S')}(-p\eta)^{\nu'}H_{\nu'}^{(1)}(-p\eta)\eqend{,}
\end{equation}
where
\begin{equation}
C^{(S')}(p) = - \kappa \frac{\sqrt{\pi\eta_0}}{2(n-2)\epsilon_0(p\eta_0)^{\nu'}}\eqend{.}
\end{equation}
The functions $Q_p^{(1)}(\eta)$, and $Q_p^{(2)}(\eta)$, and the constants
$A^{(S)}(p)$, and $B^{(S)}(p)$, are given by replacing $\nu$ by $\nu'$ and $\epsilon$ by
$\epsilon_0$ in \eref{Q1-p-eta}-\eref{BS-p}.

\section{Discussion}\label{discussion}

In this paper we studied the nature of IR divergences in the free two-point functions for the tensor perturbations in
general FLRW
spacetime, and the scalar perturbations in single-field inflation.  These IR divergences occur because, for small $p$, the mode functions
behave like $p^{-\nu}$ with \(\nu \geqslant (n-1)/2\), where \(p\) is the momentum, in \(n\) dimensions.  We pointed out that global shear transformations and dilation can increase the power by \(1\), i.e.\ from
\(p^{-\nu}\) to \(p^{-\nu+1}\), and showed that in fact there are large gauge transformations which
increase the power of \(p\) in the IR limit of the mode functions by \(2\).  This implies that
the two-point functions for the
tensor and scalar perturbations can be made IR finite by large gauge transformations
in a larger set of FLRW spacetimes (for the scale-invariant vacuum state) than previously thought.
Our focus was on the slow-roll inflation, but the reduction of the power of \(p\) in the IR in the \(p\)-integration is
valid for any potential $V(\phi)$ including those leading to bouncing cosmologies~\cite{Brandenberger:2016vhg}.

Our findings are consistent with the fact that the graviton and inflaton fields smeared in a
gauge-invariant manner are equivalent to the linearised Weyl tensor and another tensor whose \(p\)-dependence
is less singular than the original fields by a factor of \(p^2\)~\cite{Quant2,InflationGTpre}
(see also \cite{Canepa:2017cdr}).  This is because the
latter work indicates that the terms of order $p^0$, as well as those of order $p$, are of pure-gauge form, and this is
what we have verified.

Unlike the global shear transformations and dilation, we have not found a simple geometric interpretation for
the large gauge transformations that gauge away the terms of order $p$ in the mode functions,
which is an extension of  the global shear transformations and dilation.
It would be interesting to find a geometric picture of these  gauge transformations. It is interesting to note
in this context that the vectors $\xi_i$ specifying these gauge transformations are hypersurface orthogonal.

It would not be straightforward to incorporate interactions in the method of gauging away IR divergences presented
in this paper, for example, to discuss three-point functions relevant to non-Gaussianities.  
The obvious drawbacks are its non-locality and lack of manifest translation invariance.   It would be interesting
to investigate whether these difficulties could be overcome to construct perturbation theory for inflationary models that were
manifestly IR finite.

\section*{Acknowledgements}
We would like to thank Markus Fr\"{o}b for helpful discussion and a critical reading of the manuscript.  
The work of N.\ R.\ was supported in part by a studentship from the Engineering and Physical Sciences Research Council 
(EPSRC).

\appendix

\section{Derivation of the components of \(h_{\mu \nu}\)}\label{Appendix-A}


We derive equations \eref{hzerozero}-\eref{hkl} in this appendix.
We start from the expressions for the components of the metric perturbation given in terms of quantities that are
invariant under \emph{local} gauge transformations. These are given by equations \eref{zerozeroinv}-\eref{psiinv}.
Using the gauge condition \eref{scalarvanishgaugetime}, and the definition of the Sasaki-Mukhanov variable given in
equation \eref{sasakimukhanov}, we find
\begin{eqnarray}
X_0 = \frac{\Psi}{\phi'} = \frac{1}{2Ha} \left( Q + \Sigma \right) \eqend{.} \label{A-X0}
\end{eqnarray}
Taking the derivative of this with respect to \(\eta\), we find
\begin{eqnarray}
X_0 \p = \frac{1}{2Ha} \left( Q + \Sigma \right) \p + \frac{1}{2} (\epsilon-1) \left( Q + \Sigma \right) \eqend{.}
\label{A-X0prime}
\end{eqnarray}
Recall that \( X_k = 0\) for all $k$ [see \eref{scalarvanishgaugespace}].

Let us first consider the \(h_{00}\) component. Equation \eref{zerozeroinv} (given here again for convenience) is
\begin{eqnarray}
h_{00} = S + 2 X'_0 + 2 H a X_0 \eqend{.}
\end{eqnarray}
Using \eref{A-X0} and \eref{A-X0prime} and the equation of motion for the gauge invariant scalars, \(S\) and
\(\Sigma\), given by \eref{SEoM} and \eref{SigmaEoM}, we find
\begin{eqnarray}
h_{00} &= & - \epsilon Ha (n-3) \laplace ^{-1} Q \p + \frac{1}{Ha} \left( Q - \epsilon Ha \laplace ^{-1} Q \p \right) \p \nonumber \\
&& + \epsilon \left( Q - \epsilon Ha \laplace ^{-1} Q \p \right) \eqend{,} \\
&= &\frac{1}{Ha} Q \p - \laplace^{-1} \epsilon \left[ Q^{\prime \prime} + \left( n-2 + 2\delta\right) Ha  Q \p - \laplace Q \right] \eqend{,} \\
&= & \frac{1}{Ha} Q \p \eqend{,}
\end{eqnarray}
where the quantity in the square brackets vanishes because of the equation of motion~\eref{Q-fieldeq} for \(Q\).
The component \(h_{0k}\) is
\begin{eqnarray}
h_{0k} &= V_k + X'_k + \partial_k X_0 \eqend{,} \\
&= \frac{1}{2} \partial_k \left( \frac{1}{H a} Q - \epsilon \laplace^{-1} Q'\right) \eqend{,}
\end{eqnarray}
where the last line follows from the gauge conditions \eref{scalarvanishgaugetime} and
\eref{scalarvanishgaugespace}, and equations of motion \eref{VEoM}-\eref{SigmaEoM}.
The expression for \(h_{kl}\) also readily follows from these conditions:
\begin{eqnarray}
h_{kl} &= H_{kl} + \delta_{kl} \left[ \Sigma - (\Sigma + Q )\right] \eqend{,} \\
&= H_{kl} - \delta_{kl} Q \eqend{.}
\end{eqnarray}

\section{Proof that \( F_p^{(1)}(\eta) - F_0^{(1)}(\eta) \) is of order \( p^2\) for constant $\epsilon$}
\label{Appendix-B}

In this appendix we demonstrate that, in the constant-$\epsilon$ case, $F_p^{(1)}(\eta) - F_0^{(1)}(\eta)$,
where $F_p^{(1)}(\eta)$ is defined by
\eref{F1_p}, is of order $p^2$ if $\nu \geqslant (n-1)/2$, i.e.\ if the original two-point function is IR divergent.
The proof that \( Q_p^{(1)}(\eta) - Q_0^{(1)}(\eta) = \mathcal{O}(p^2)$ for constant
$\epsilon$ or for slow-roll inflation is almost identical.

Define
\begin{equation}
G_\nu (z) : = z^\nu Y_\nu(z)\eqend{.}
\end{equation}
All we need to show is that $G_\nu (z) - G_\nu (0) = \mathcal{O}(z^2)$ for small $z$.  If $\nu$ is not an integer, then
we have
\begin{equation}
z^\nu Y_\nu(z) =  - \frac{1}{\sin(\pi\nu)} z^{\nu}J_{-\nu}(z) + \cot (\pi \nu) z^{\nu}J_\nu(z)\eqend{,} \label{zYnu}
\end{equation}
where
\begin{equation}
J_\nu(z) = \sum_{k=0}^\infty \frac{(-1)^k}{k!\Gamma(k+1+\nu)}z^{2k+\nu}\eqend{.}
\end{equation}
The second term in \eref{zYnu} is $\mathcal{O}(z^{2\nu})$.  The original two-point function is IR divergent if
$\nu \geqslant (n-1)/2$ because the power of $p$ in the integral for small $p$ is $p^{-2\nu}$ whereas the integration
measure is $\mathrm{d}p\,p^{n-2}$.  Thus, for $n\geqslant 3$, if the original two-point function is IR divergent, then
the second term in \eref{zYnu} is $\mathcal{O}(z^2)$ or higher, which can be neglected.
Then since
\begin{equation}
z^{\nu}J_{-\nu}(z) = \sum_{k=0}^\infty \frac{(-1)^k}{k! \Gamma(k+1-\nu)}z^{2k}\eqend{,}
\end{equation}
it is clear that $G_\mu(z)-G_\mu(0) = \mathcal{O}(z^2)$.

If $\nu=N$ is a positive integer, then
\begin{eqnarray}
z^{-N}Y_N (z) =& \frac{2}{\pi} \left[ \log \left( \frac{z}{2} \right) + \gamma \right] z^NJ_N (z)
- \frac{1}{\pi} \sum_{k=0}^{N-1} \left[ (N-k-1)! \left( \frac{z}{2} \right)^{2k} \right] \nonumber \\
&- \frac{1}{\pi} \sum_{k=0}^{\infty} \left[ (-1)^k \left( \frac{\phi(k) + \phi(N+k) }{k! (N+k)!} \right) \left( \frac{z}{2}
\right)^{2k +2N} \right] \eqend{,}
\end{eqnarray}
where \(\phi (p) = \sum_{n=1}^p \frac{1}{n}\). Therefore, as the power of \(z\) increases in increments of 2 in the first
sum, the same argument as above holds and we conclude that $G_N(z)-G_N(0) = \mathcal{O}(z^2)$.

\section*{References}

\bibliographystyle{unsrt}

\bibliography{IR_cosmology_papers}
\end{document}